\documentclass[onecolumn,english,aps,pra,superscriptaddress,letterpaper,preprint]{revtex4-1}

\pdfoutput=1

\usepackage[pdftex]{graphicx}
 
\usepackage[a4paper, margin=0.9in]{geometry}
\usepackage{amsmath,amssymb}

\usepackage{siunitx}
\usepackage[T1]{fontenc}
\usepackage[utf8]{inputenc}
\usepackage{float}
\usepackage{adjustbox}

\begin{document}
\title{\bf{Nonlinear carrier dynamics in silicon nano-waveguides}}


\author{I.~Aldaya} \email{These authors contributed equally to this work}
\affiliation{Gleb Wataghin Physics Institute, University of Campinas, Campinas, SP, Brazil}
\author{A.~Gil-Molina} \email{These authors contributed equally to this work}
\affiliation{Gleb Wataghin Physics Institute, University of Campinas, Campinas, SP, Brazil}
\affiliation{School of Electrical and Computer Engineering, University of Campinas, Campinas, SP, Brazil}
\author{J.L.~Pita}
\affiliation{School of Electrical and Computer Engineering, University of Campinas, Campinas, SP, Brazil}
\author{L.H.~Gabrielli}
\affiliation{School of Electrical and Computer Engineering, University of Campinas, Campinas, SP, Brazil}
\author{H.L.~Fragnito}
\affiliation{Gleb Wataghin Physics Institute, University of Campinas, Campinas, SP, Brazil}
\affiliation{MackGraphe - Graphene and Nanomaterials Research,
Mackenzie Presbyterian University, S\~{a}o Paulo, SP, Brazil}
\author{P.~Dainese} \email{Corresponding author: dainese@ifi.unicamp.br}
\affiliation{Gleb Wataghin Physics Institute, University of Campinas, Campinas, SP, Brazil}


\begin{abstract}
Carrier recombination dynamics in strip silicon nano-waveguides is analyzed through time-resolved pump-and-probe experiments, revealing a complex recombination dynamics at densities ranging from ${10^{14}}$ to  ${10^{17}}\,$cm$^{{-3}}$. Our results show that the carrier lifetime varies as recombination evolves, with faster decay rates at the initial stages (with lifetime of ${\sim 800}\,$ps), and much slower lifetimes at later stages (up to ${\sim 300}\,$ns). We also observe experimentally the effect of trapping, manifesting as a  decay curve highly dependent on the initial carrier density. We further demonstrate that operating at high carrier density can lead to faster recombination rates. Finally, we present a theoretical discussion based on trap-assisted recombination statistics applied to nano-waveguides. Our results can impact the dynamics of several nonlinear nanophotonic devices in which free-carriers play a critical role, and open further opportunities to enhance the performance of all-optical silicon-based devices based on carrier recombination engineering.
\end{abstract}
\maketitle
\email

\section{Introduction}

Free-carrier effects have a critical role in future silicon photonic circuits~\cite{leuthold2010,liang2004,yin2007,blanco2014}.
Fundamentally, both the dispersion and attenuation of optical modes in waveguides and cavities are modified in the presence of excess electron-hole pairs in the silicon core region, effects referred respectively as Free-Carrier Dispersion ({\small FCD}) and Free-Carrier Absorption ({\small FCA})~\cite{Soref1987,lin2007}.
These two basic phenomena have been extensively explored in a variety of silicon-based photonic devices and applications.
For example, carriers injected externally through a p-i-n structure in ring-resonators or in integrated Mach-Zenhder interferometers have been used to build fast optical modulators based on {\small FCD}-induced phase-shift~\cite{Baba2013,Xu1012}.
Other devices based on {\small FCA} have also been demonstrated such as waveguide based optical attenuators with externally injected carriers~\cite{Park1010}.

Even in the absence of external injection, excess carriers can be generated optically due to silicon's relatively high Two-Photon Absorption ({\small TPA}) coefficient at the \SI{1550}{nm} telecommunication wavelength ($\sim$\SI{0.7}{cm/GW})~\cite{bristow2007}.
All-optical modulation has been achieved using {\small TPA}-generated free carriers by a high power control pump pulse~\cite{almeida2004}.
Given silicon's high refractive index, in sub-micron structures the optical mode is tightly confined and nonlinear effects (such as {\small TPA}) appear at relatively low power.
Several nonlinear phenomena and applications are therefore impacted by {\small TPA}-induced free-carriers \cite{yin2007,yin2007_1,liang2004}.
For example, free-carriers impact the stability of soliton propagation and self-breathing phenomena in silicon waveguides~\cite{blanco2014}, give rise to {\small FCD}-induced soliton self-frequency shift, limit the efficiency of parametric and Raman amplification~\cite{liang2004}, improve the coherence while reducing the efficiency of supercontinuum generation~\cite{leo2015}, and limit the gain obtained in stimulated Brillouin scattering~\cite{shin2013}.
Analogously, there are also a number of nonlinear phenomena impacted by free-carriers in micro-cavities~\cite{johnson2006self, pernice2010}.

{\small FCD} and {\small FCA} are not the only effects caused by excess carriers.
Because silicon is a material with indirect bandgap, excess carriers recombine dominantly through a phonon-assisted process, which ultimately gives rise to an increase in the device temperature.
This in turn modifies the refractive index through the thermo-optic effect.
In this context, self-oscillation in micro-cavities is an interesting example~\cite{johnson2006self,carmon2004,pernice2010}: first, the cavity resonance shifts due to {\small FCD}, and second, an opposite shift arises due to the temperature increase as a result of carrier recombination.

In the applications discussed above, any time domain analysis must take into account the dynamics of carrier generation, spatial diffusion and recombination.
Optical generation is generally assumed instantaneous relative to the time scale in most photonic applications.
Once a certain spatial distribution of carriers is created (e.g.\ following the square of the intensity profile in {\small TPA} generation), diffusion takes place.
Obviously the rate at which carriers diffuse and the evolution of the spatial charge profile depends on the particular geometrical structure as well as on carrier mobility.
For example, carrier diffusion has been extensively modeled in photonic crystal cavities~\cite{tanabe2008, nozaki2010sub} as well as in rib-waveguide structures~\cite{dimitropoulos2005}.
In those, carriers can diffuse out of the region in which the optical mode is confined.
As a consequence, their impact on the optical mode (through {\small FCD} and {\small FCA}) ceases even before these carriers have recombined back to the valence band, simply because they have left the modal region.
This is not necessarily the case in silicon strip waveguides as the silicon core is completely surrounded by a dielectric material.
The spatial distribution within the silicon core can evolve due to diffusion, but carriers no longer leave the modal region.
In this case, recombination determines the rate at which free-carriers cease to impact the optical mode.

In most photonic applications, carrier recombination is treated using an exponential time decay curve, generally characterized by a single lifetime.
Although the single-exponential decay is justified under certain conditions (i.e.\ minority-dominated carrier lifetime), generally speaking it is well known that recombination processes are strictly not single-exponential~\cite{blakemore2002}.
In silicon, band-to-band radiative recombination is generally neglected due to its indirect bandgap and the main recombination mechanisms are: (i) Auger recombination, which is significant only at high carrier densities (above $10^{18}$~cm$^{-3}$)~\cite{sze2006physics}, and (ii) trap-assisted recombination, dominant in most cases~\cite{shockley1952,schroder1997}.
In this paper, we explore the recombination of carriers in a silicon strip waveguide, under conditions that allow clear observation of complex recombination dynamics, particularly non-exponential decay.
Using a pump and probe technique, we characterized the carrier dynamics for different excitation pulse powers and durations.
Our results reveal faster decay rates at initial stages of recombination and slower ones at later stages. We observe experimentally the effect of trapping, leading to memory in the decay dynamics and we also demonstrate that operation at high carrier density leads to faster recombination rates.
This paper is organized as follows: in Section~\ref{s:experiment} we describe our samples and our experimental methods; in Section~\ref{s:results} we present our results, and discuss their implications to all-optical switching.
In Section~\ref{s:discuss} we present a discussion of the results in terms of trap-assisted recombination and draw our conclusions in Section~\ref{s:conclusion}.


\section{Samples and experimental setup}
\label{s:experiment}

\begin{figure}
\centering
\includegraphics[clip=true]{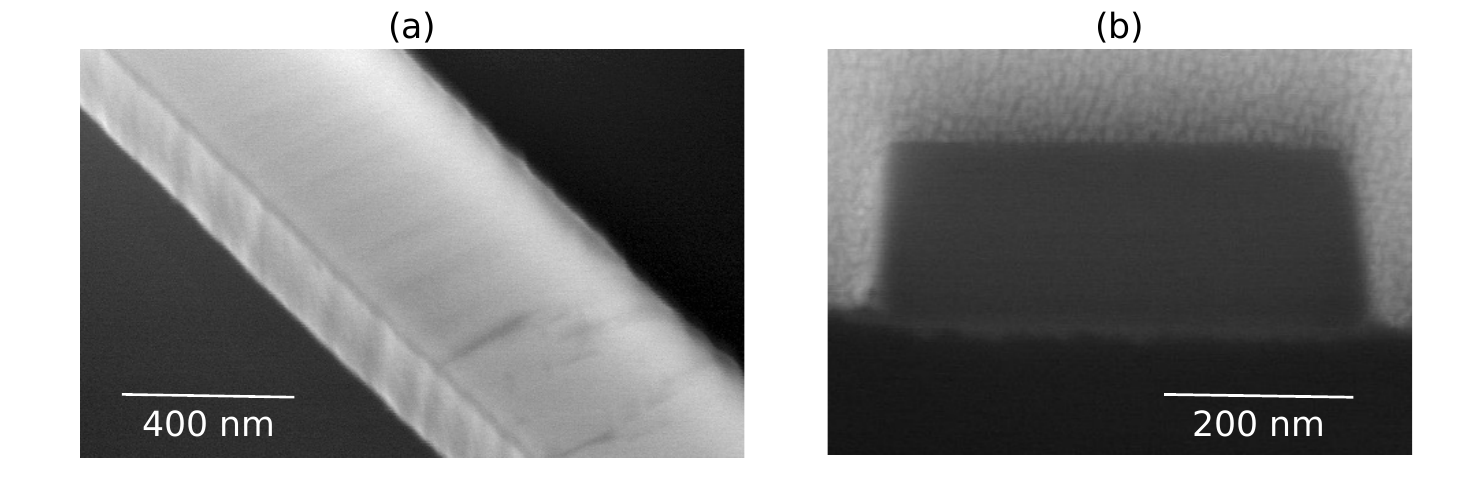}
\caption{Scanning electron microscope images for an unclad silicon strip waveguide of $\SI{450}{nm} \times \SI{220}{nm}$, similar to the one used in our experiments: (a)~perspective and (b)~cross-section views.}
\label{Fig:wg_sem}
\end{figure}

We analyzed silicon on insulator waveguides with a cross-section of $\SI{450}{nm} \times \SI{220}{nm}$ and length \SI{5.9}{mm}.
All samples had silicon dioxide cladding and were fabricated at the imec/Europractice facility.
A scanning electron microscope image of an unclad sample (before oxide deposition) can be seen in Fig.~\ref{Fig:wg_sem}.
Light was coupled in and out of the waveguides using grating couplers.
Coupling and propagation losses were evaluated at \SI{3.1}{dB} and \SI{1.4}{dB/cm} through linear regression of the measurements of three samples with different lengths (2.4, 5.9, and \SI{30}{mm}) under low input power.

\begin{figure}
\includegraphics[clip=true, scale=0.9]{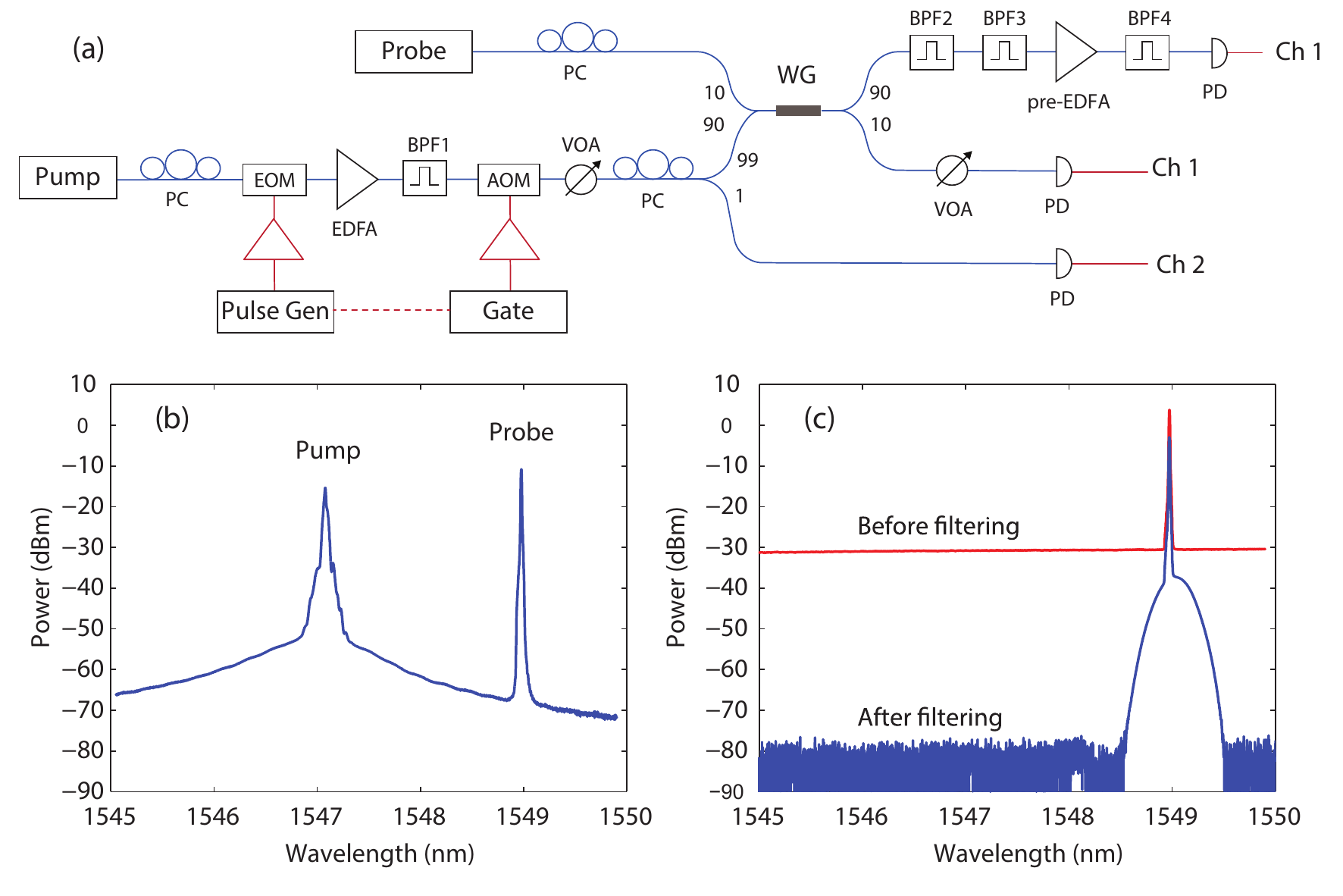}
\caption{(a)~Experimental setup employed to characterize the free-carrier lifetime in {\small SOI} strip waveguides.
{\small WG}: waveguide under test.
Other acronyms are defined within the text.
(b)~Power spectrum at the input of {\small WG}.
(c)~Power spectra before and after \small{BPF4}.
}
\label{Fig:Setup}
\end{figure}

Figure~\ref{Fig:Setup}a shows the pump and probe experimental setup employed to characterize the free-carrier dynamics in the waveguide under test.
The pulsed pump was synthesized by externally modulating a continuous wave ({\small CW}) laser operating at \SI{1547}{nm} (with \SI{20}{mW} optical power) using a Mach-Zehnder electro-optical modulator ({\small EOM}).
The {\small EOM} was driven by a train of pulses with \SI{500}{kHz} repetition rate and pulse duration ranging from \SI{130}{ps} to \SI{20}{ns}.
The {\small EOM} used has \SI{20}{GHz} bandwidth and more than \SI{30}{dB} extinction ratio.
The pump signal was then amplified using an erbium-doped fiber amplifier ({\small EDFA}).
Special care was taken to avoid generation of free-carriers outside the pump pulse window.
First, the {\small EOM} bias voltage was set for maximum peak pump power at the output of the {\small EDFA}, minimizing any remaining {\small CW} level on the pump.
Second, the output of the {\small EDFA} was filtered using an optical bandpass filter ({\small BPF}) to reduce the out-of-band {\small ASE} and an acousto-optic modulator {\small(AOM)} operating as an optical gate filtered out any remaining {\small CW} components outside the pulse window (either from the pump or {\small ASE}).
The {\small AOM} was driven with a \SI{20}{ns} gate pulse duration, and has an extinction ratio larger than \SI{50}{dB}.
The pump power was controlled with a variable optical attenuator ({\small VOA}) and a 1\% fraction was derived and monitored in a scope for stability and power control.

The pump was then combined with the {\small CW} probe (operating at \SI{1549}{nm}) using a 10-90 optical coupler and then injected into the waveguide. The probe power was set to \num{-12}~dBm in the waveguide.
Figure~\ref{Fig:Setup}b shows the spectrum measured at the input, before coupling into the waveguide.
The {\small ASE} power spectral level is below \SI{-60}{dBm} in a \SI{0.01}{nm} resolution bandwidth.
Polarization controllers ({\small PC}'s) were used on both pump and probe arms to optimize coupling through the grating, which was designed for transverse-electric operation.
After the waveguide, 10\% of the output signal was monitored in a scope and 90\% was filtered through two cascaded {\small BPF}s to remove the pump (with more than 40 and \SI{50}{dB} of rejection ratios).
Before detection, the probe was amplified using a low-noise pre-{\small EDFA} and then filtered with a {\small BPF} to remove out-of-band {\small ASE} (with more than \SI{50}{dB} of rejection ratio).
Figure~\ref{Fig:Setup}c shows the spectra before and after {\small BPF4} identified in Fig.~\ref{Fig:Setup}a.
A wide-bandwidth oscilloscope was used to capture the signals.
The input to Channel 1 (an optical input with a built-in \SI{28}{GHz} photodetector, {\small PD}) was switched to measure either the filtered probe or the output pump pulses, whereas in Channel 2 (electrical input), the output of an external {\small PD} of \SI{20}{GHz} bandwidth was captured in order to monitor the input pump pulses.

\section{Results}
\label{s:results}

\begin{figure}
\includegraphics[clip=true, scale =0.83]{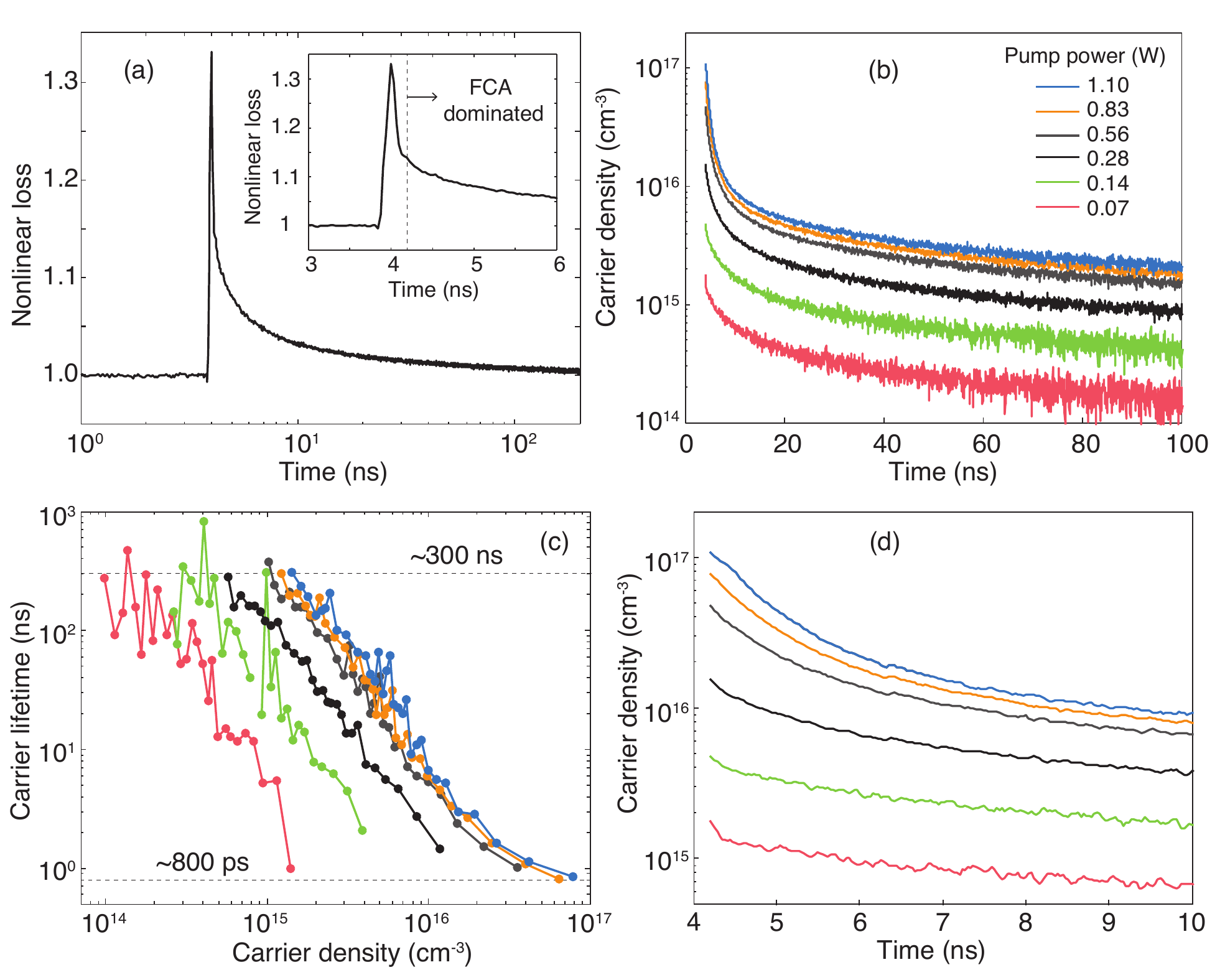}
\caption{Analysis of carrier density dynamics using \SI{130}{ps} pump pulses.
(a)~Nonlinear loss as a function of time for \SI{0.28}{W} pump power.
The inset shows a zoom of the normalized nonlinear loss around the pump pulse in linear time scale.
(b)~Carrier density as a function of time for different pump peak powers.
(c)~Recombination lifetime as a function of carrier density for different pump peak powers.
(d)~Detail of the carrier density as a function of time for different pump peak powers in the first few nanoseconds.}
\label{Fig:CarrierDynamics}
\end{figure}

From the detected output probe signal, the nonlinear loss is obtained from the ratio between the detected signals for pump off and on (which therefore excludes linear losses).
Since we are using wideband photodetectors in order to observe rapid lifetimes, several measurements were averaged to reduce high-frequency noise in the nonlinear loss curves (25 and 400 for the highest and lowest power, respectively).
An example of the detected nonlinear loss is shown in Fig.~\ref{Fig:CarrierDynamics}a for a \SI{130}{ps} pump pulse with \SI{280}{mW} peak power in a waveguide of \SI{5.9}{mm}.
In the inset, two regimes can be identified: within the pump pulse duration, the nonlinear loss is dominated by instantaneous non-degenerate {\small TPA}, while after the pulse the nonlinear loss arises solely from {\small FCA}~\cite{fishman2011, zhang2015non, meitzner2013time}.
For waveguides with larger cross sections or structures that allow carriers to leave the modal region (e.g.\ photonic crystal cavities), an intermediate stage where diffusion plays a role has been reported~\cite{liu2010,tanabe2008}.
However, as already mentioned, the nanowire structure investigated here confines the excess carriers to the core region, not allowing them to diffuse out.
Moreover, for the small dimensions of the waveguide cross-section explored in this paper, a non-uniform initial carrier distribution (just after the pulsed pump generation has ceased) diffuses throughout the entire core and becomes uniform in a relatively short period of time.
In Supplementary Material A, an initial Gaussian distribution is shown to become uniform in approximately \SI{5}{ps} for electrons and \SI{15}{ps} for holes, which is too short to be identified in Fig.~\ref{Fig:CarrierDynamics}a.
In the {\small FCA}-dominated regime, the nonlinear loss is determined by the accumulated {\small FCA} along the waveguide length:
\begin{equation}
L_{FCA}(t)=\exp\!\left[\alpha_r\int_0^L N(z,t)\,{\rm d}z\right]=\exp\left[\alpha_r\bar N(t) L\right],
\label{eq:TFCA}
\end{equation}
%
where $\alpha_r = \SI{1.45e-21}{m^2}$ is the {\small FCA} cross-section in silicon at \SI{1550}{nm}~\cite{lin2007, Soref1987}, $N(z,t)$ is the carrier density at a certain position $z$ along the waveguide at an instant of time $t$, $L$ is the waveguide total length.
Here $\bar{N}(t)$ is the average of the carrier density along the waveguide length.
By inverting Equation~\ref{eq:TFCA}, we can extract the time-resolved carrier density average $\bar{N}(t)$ from the measured nonlinear loss $L_{FCA}$.
As shown in detail in Supplementary Material B, for the power levels and waveguide length explored in this paper, $\bar{N}(t)$ approximates reasonably well $N(z,t)$.
However, large deviations occur as pump power or waveguide length increase.
From now on, we refer to $\bar{N}(t)$ as simply carrier density (not specifying it is the averaged value).

Using the nonlinear loss from Fig.~\ref{Fig:CarrierDynamics}a in the {\small FCA}-dominated regime (i.e. for times after \SI{4.2}{ns}), the obtained carrier density as a function of time is shown in Fig.~\ref{Fig:CarrierDynamics}b (black curve).
The result clearly shows that the free carrier density does not decay following a simple exponential curve, as it is not a straight line in logarithmic scale.
We repeated these measurements for pump power levels ranging from \SI{0.07}{W} to \SI{1.1}{W}, and the results are also shown in Fig.~\ref{Fig:CarrierDynamics}b.
A nonlinear decay behavior is observed in all curves, with faster decay rates at the beginning and slowing down as recombination progresses and the density falls.
At the final stages of recombination, all curves approach the same slope, however they differ significantly at early stages.

Several remarks can be made from this result.
First, as already stated, the instantaneous carrier lifetime varies as the recombination evolves.
This can be assessed quantitatively by numerically computing $-\frac{\bar{N}}{{\rm d}\bar{N}/{\rm d}t}$ (see Supplementary Material C for details in the numerical slope computation).
The results are shown in Fig.~\ref{Fig:CarrierDynamics}c for the same power levels from Fig.~\ref{Fig:CarrierDynamics}b.
In all curves, the instantaneous lifetime varies from a slow-limit of $\sim$\SI{300}{ns} to a fast-limit of $\sim$\SI{800}{ps}.
This represents more than 2 orders of magnitude reduction in the instantaneous lifetime as the density decays over almost 3 orders of magnitude.
As already mentioned, the slow decay limit can be seen directly from Fig.~\ref{Fig:CarrierDynamics}b (here, it is important to ensure that the probe power is low enough not to impact the slow decay lifetime---see Supplementary Material D for details).
The fast limit can be appreciated in Fig.~\ref{Fig:CarrierDynamics}d, which shows the density decay at the first few nanoseconds.

A second remarkable observation is that the lifetime is not simply a function of the carrier density.
This is seen directly in Fig.~\ref{Fig:CarrierDynamics}c, in which each curve has a different instantaneous lifetime for the same value of excess carrier density.
The same conclusion can be drawn directly from the decay curves in Fig.~\ref{Fig:CarrierDynamics}b, where at a given density value, the decay trajectory, i.e.\ $\bar{N}(t)$, is different for different initial values $\bar{N}(0)$, a form of memory in the decay dynamics.
In Section~\ref{s:discuss}, we provide an explanation to this observation in terms of carrier trapping at the recombination centers, which leads to electrons and holes following different decay curves.

A final important observation in Fig.~\ref{Fig:CarrierDynamics}c is regarding the curves corresponding to the highest peak power levels.
Note that the decay lifetime remains fast at around a few nanoseconds for a wide range of density between $10^{16}$ and $10^{17}\,$cm$^{-3}$.
In contrast, the lifetime for the lowest peak power curve increases from about \SI{1}{ns} to over \SI{100}{ns} in just one order of magnitude change in density, from $10^{15}$ and $10^{14}\,$cm$^{-3}$.
This observation leads to the conclusion that operating at high carrier densities can be used as a strategy to obtain faster all-optical switching, as demonstrated in the next section.

To summarize this discussion, three key observations are highlighted: (i)~the carrier lifetime is faster initially and becomes slower as recombination evolves, with lifetimes ranging at least 2 orders of magnitude; (ii)~the decay curve is not well defined by simply specifying the carrier density, but depends on its initial value; and (iii)~operating at high density leads to faster decay rates for a wider density range.
This behavior is discussed in detail in Section~\ref{s:discuss}, considering the well established statistics of trap-assisted recombination process.

\begin{figure}[b]
\includegraphics[scale = 0.83]{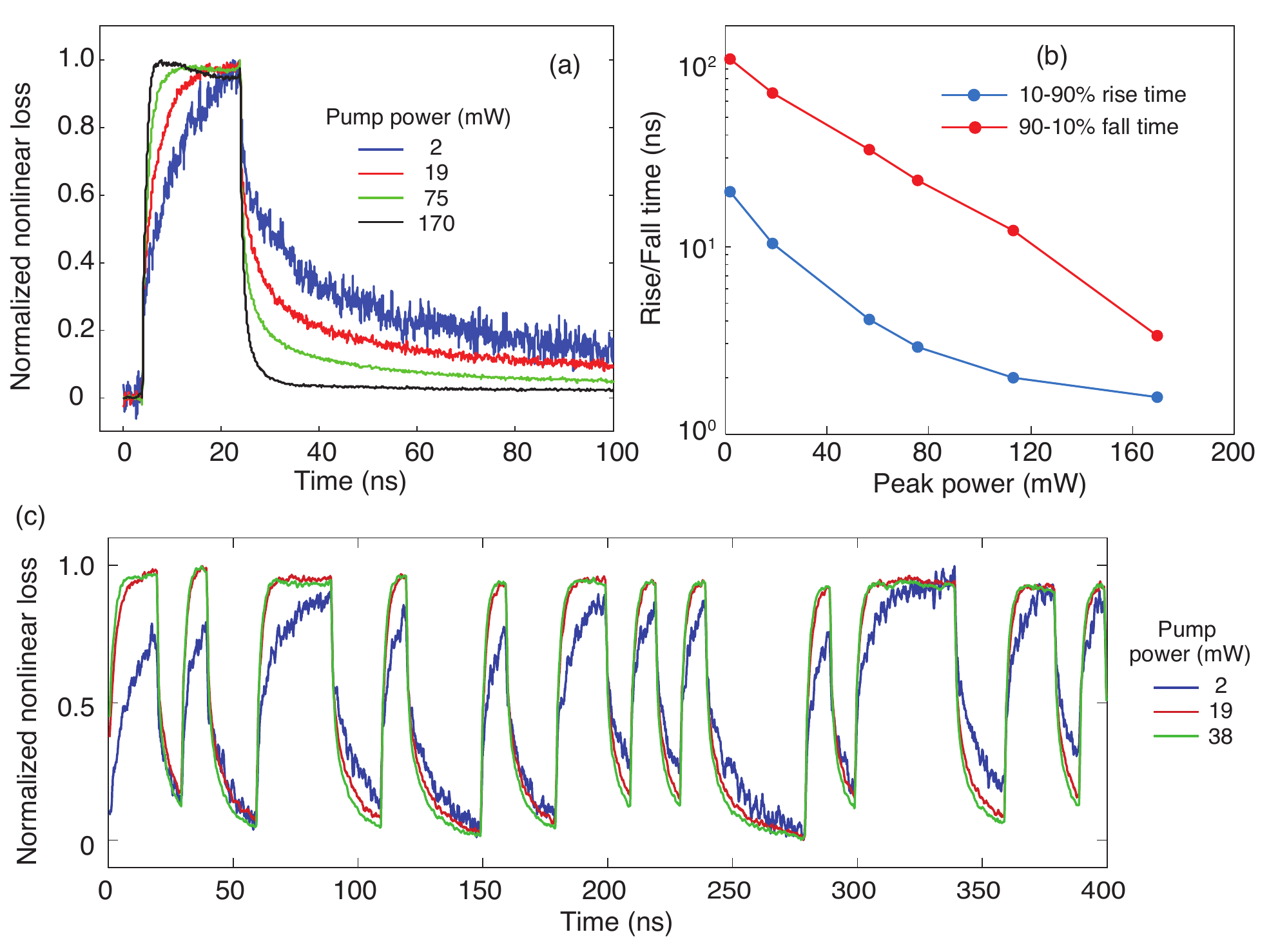}
\caption{Analysis of the carrier recombination dynamics using \SI{20}{ns} pulses.
(a)~Normalized nonlinear loss for pump power levels of 2, 19, 75 and \SI{170}{mW}.
(b)~Rise and fall times for different pump peak powers.
(c)~Nonlinear loss for a pseudo-random sequence of \SI{10}{ns} pulses for three power levels of 2, 19, and \SI{38}{mW}.}
\label{Fig:LongPulses}
\end{figure}

Once the carrier recombination dynamics has been analyzed, we now demonstrate how it
is affected by the pump power level and the accumulated free-carriers of previous pump pulses. 
We used long pump pulses (10 and \SI{20}{ns}) in order to quantify not only the decay rate but also the carrier build-up dynamics.
Figure~\ref{Fig:LongPulses}a shows the normalized nonlinear losses obtained for various pump power levels ranging from $\sim$2 to $\sim$\SI{170}{mW}.
As expected from the previous dynamic analysis, higher pump power results in a faster response
not only in the decay stage but also in the build-up stage.
Figure~\ref{Fig:LongPulses}b shows the rise and fall times calculated as the period it takes for the nonlinear loss to change from 10\% to 90\% and from 90\% to 10\%, respectively.
The curves confirm a strong reduction in response
time with increased pump power.
Figure~\ref{Fig:LongPulses}c shows the nonlinear loss measured for \SI{10}{ns} pump pulses in a 40-symbols pseudo-random sequence at different power levels.
This results confirms that higher pump power results in a faster behavior.
It is interesting to note that since the instantaneous recombination rate strongly depends on the carrier dynamics, the dynamics is word-dependent.
This can be appreciated at the beginning of the sequence: because the initial carrier density in the waveguide was low, the system is relatively slow, specially for \SI{2}{mW} pump power.
As free-carriers are accumulated, the system becomes faster.
This suggests a strategy to increase the overall speed by injecting a {\small CW} pump component to offset the excess carrier density, similarly to what is explored in the context of solar cells~\cite{aberle1992impact}.

We make a final comment in this section regarding the absolute values of the lifetimes reported here.
The minimum lifetime at high carrier density is observed at around \SI{800}{ps}.
This is however not a fundamental limit.
As we discuss in detail in the next section, this value can be reduced by either increasing the density of recombination centers (for example by increasing the surface flaw density on the waveguide side and top walls), as well as reducing the waveguide dimensions, since in surface recombination the lifetime is directly proportional to the waveguide size.

\section{Discussion}
\label{s:discuss}

There are two key observations in our experiments that the present discussion elucidates: first, the non-exponential decay curve with faster rates initially and slower rates at the final stage.
Second, the fact that the shape of the transient decay curve depends on the initial condition, i.e., on the initial excess carrier concentration, a form of memory behavior.

We focus this discussion on trap-assisted recombination.
Although this is a well-established process~\cite{shockley1952,blakemore2002,aberle1992impact}, its application in silicon photonic waveguides and cavities remains unexplored to the best of our knowledge.
Through this process, an excess carrier, say an excess electron, is first captured in a trap state (referred generically here as a flaw) and eventually transferred to the valence band when an excess hole is captured by the same trap state.
In {\small TPA}, excess electrons and holes are generated in equal numbers.
However, it may occur that as recombination takes place, a significant unbalance is created between the density of excess electrons in the conduction band ($n_{e}$) and the density of excess holes in the valence band ($p_{e}$).
This unbalance simply means that a significant fraction of electrons (or holes) are trapped in the flaws for a certain period of time.
In other words, electrons and holes do not necessarily decay at the same rate due to trapping.
In our experiments, we measure absorption due to free-carriers, and cannot distinguish between free-electrons and free-holes.
In fact, the decay dynamic we observe experimentally reflects very nearly the dynamics of the sum of excess electron and hole concentrations.
The exact free-carrier absorption coefficient is given by the weighed sum of excess carriers $(8.5 n_{e} + 6.0 p_{e})\times 10^{-18}\,$cm$^{-1}$, each multiplied by its own absorption coefficient~\cite{lin2007, Soref1987}.
However, for simplicity, we discuss the behavior of the total excess density $n_{e}+p_{e}$.
As we shall see, initially one type of carrier (electron or hole) decays faster than the other due to trapping, resulting in a non-exponential decay curve for the total $n_{e}+p_{e}$.

In order for significant trapping to occur, the density of flaws ($D_{f}$) must be relatively large---at least comparable to the density of excess carriers.
If the flaw density is too small, there cannot be a significant unbalance between $n_{e}$ and $p_{e}$, because even if all flaws are filled, the total number of trapped carriers would still be small compared to the total number of free-carriers.
In the absence of trapping, excess electrons decay at the same rate as excess holes.
This regime is usually referred to as Shockley-Read-Hall ({\small SRH}) recombination~\cite{shockley1952,blakemore2002}.
Interestingly, it can also lead to non-exponential decay since lifetime for high-excess density is different from the lifetime for low-excess density.
However, as we discuss here, in the absence of trapping one cannot explain the dependence of the transient decay curve on the initial condition---clearly observed in our experiments.
In the {\small SRH} model, all transient decay curves follow the same path and a change in initial condition is simply a time shift of the decay curve.
In other words, in the {\small SRH} model the lifetime is a well defined function of carrier density, which is not in agreement with our experiments.
We therefore attribute our observation to the presence of trapping.


Flaws can be located throughout the volume of the silicon core or at the interface between silicon and silicon-dioxide.
The latter is usually assumed to dominate in nano-waveguide (i.e., with large surface-to-volume ratio)~\cite{dimitropoulos2005}.
Based on a simple geometrical argument, it is possible to show that if surface flaws dominate, then the smaller the waveguide is, the more likely significant trapping is to occur.
This can be seen as follows: if the recombination is dominated by volume flaws, then the fraction of flaws that are occupied (i.e., captured an electron or a hole for donor-like or acceptor-like flaws, respectively) is simply $(p_{e}-n_{e})/D_{f}$.
This simply states that any unbalance in excess electrons and holes must be trapped in the flaws so that charge neutrality is maintained.
If $D_{f}$ is large, then one can have a significant unbalance ($p_{e}-n_{e}$), obviously limited to when occupancy reaches 100\% (saturation of flaws).
Note that $p_{e}$, $n_{e}$ and $D_{f}$ are all volume density and therefore scale together as the waveguide dimensions change.
For flaws located on the surface, this same line of argument leads to a surface-to-volume ratio dependency.
The quantity $(p_{e}-n_{e}) A_{c} L$ represents the total number of carriers that must be trapped in the flaws in order to maintain charge neutrality (here $A_{c} = W H$ is the core area and $L$ is the waveguide length).
If we call $D_{s}$ the density of flaws per unit area and $A_s = 2 H L$ the sidewall area (assuming most flaws to be located on the sidewalls), then the fraction of flaws that are now occupied by excess carriers is simply $ (p_{e}-n_{e})\frac{W}{2 D_{s}}$.
This expression states that the smaller the waveguide width $W$ is, the higher the unbalance can be, even if one did not increase the flaw surface density $D_{s}$.
In here perhaps lies the explanation to why this nonlinear decay dynamics becomes readily evident in our nano-waveguide samples.
In the discussion that follows we use the symbol $D_{f}$ to represent the flaw density.
If applied to bulk flaws, $D_{f}$ gives directly the flaw volume density.
However, if one applies the theory below to surface recombination, then $D_{f} = 2 D_{s}/W$ should be used.
In what follows, a large flaw density $D_{f}$ should always be interpreted as either a truly increased surface density $D_{s}$ or simply a reduction in the waveguide width $W$.

The decay dynamics in the case of a single flaw energy level is governed by the following nonlinear equations that couple together electron and hole excess densities~\cite{blakemore2002}:
\begin{align}
\frac{{\rm d}x}{{\rm d}\tau} &= g -\frac{(x-y)[x+a(1+b)]}{D} - \frac{x}{1+b},
\label{eq:dxdt} \\
\frac{{\rm d}y}{{\rm d}\tau} &= g -\frac{\gamma (y-x)(y+1+b)}{D} - \frac{\gamma b y}{1+b},
\label{eq:dydt}
\end{align}
where we have used normalized variables defined in Table~\ref{tableparam}.
Here $g$ is the normalized generation rate, assumed to be the same for excess electrons and holes.
Also, $x$ and $y$ are the normalized excess electron and hole densities while $D$ is the normalized flaw density.

To illustrate the transient decay (i.e., after generation has ceased), we solved equations~\ref{eq:dxdt} and~\ref{eq:dydt} numerically assuming an impulse excitation, so that $x(0) = y(0)$ (as in a short pulse {\small TPA}-generated carriers), thus $n(0) = x(0) + y(0)$ is the normalized total carrier density at the beginning of the transient decay.
We chose a donor-like flaw with electron capture cross-section $\sigma_n$ larger than hole capture cross-section $\sigma_p$, which is typical for SiO$_{2}$-Si interface~\cite{aberle1992impact}, with ratio $\sigma_n = 8 \sigma_p$, resulting in $\gamma = 0.1$ for the values of electron and hole thermal velocities given in Table~\ref{tableparam}.
We also assumed that the flaw energy level is located near the middle of the bandgap (at \SI{0.5}{eV} above the valence band).
Since our sample is a $\sim$\SI{10}{\ohm\cm} p-type semiconductor, the initial flaw occupancy is $b/(1+b)=\num{4e-6}$, which means essentially all flaws are unoccupied and ready to capture an electron.
Other parameters used in the simulations are given in Table~\ref{tableparam}.

\begin{figure}
\centering
\includegraphics[scale=1]{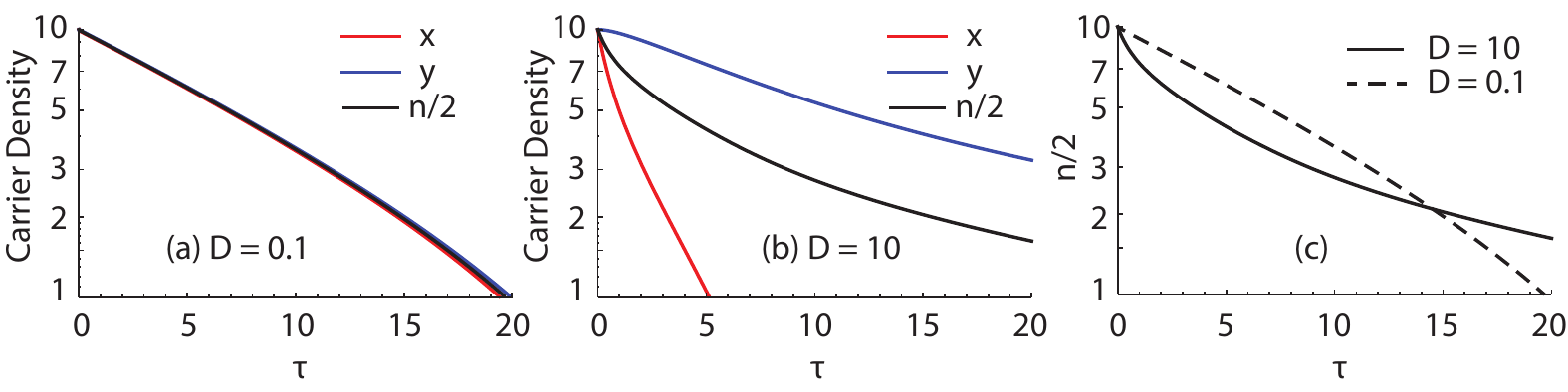}
\caption{Normalized carrier density transient decay for (a)~small, and (b)~large flaw density.
(c)~Total normalized carrier density $n$ shown for both normalized flaw densities $D = 0.1$ and $D=10$.
In all figures, we plotted $n/2$ for better visualization.}
\label{fig:transdecay}
\end{figure}

Figures~\ref{fig:transdecay}a and ~\ref{fig:transdecay}b show the transient decay for small ($D=0.1$) and large ($D=10$) normalized flaw densities, respectively.
In red and blue are the decay for electrons ($x$) and holes ($y$) respectively, while in black we show the decay for the sum of electron and hole densities $n(\tau) = x(\tau) + y(\tau)$ (divided by two for better visualization).
Clearly, for small flaw density, both excess electrons and excess holes decay at the same rate (i.e., no significant trapping occurs).
On the other hand, for large flaw density, electrons decay faster than holes.
It is interesting to note that in this initial period most electrons are being trapped, and not immediately returning to the valence band.
Simultaneously, the excess holes initially find very few filled flaws to be captured, and therefore recombine slowly.
As time goes by, a significant fraction of the flaws become filled and the electron capture rate decreases, while the hole capture rates increases.
After long enough time (not shown here) the decay lifetime reaches a steady state with equal values for both electrons and holes.

As a result of this dynamics, the decay curve for the total (normalized) density $n$ also becomes very nonlinear.
Initially, it decays following the fast electron decay.
At longer times, the density of electrons becomes much smaller that the one of holes, and then $n$ decays following the slower hole transient decay.
In Fig.~\ref{fig:transdecay}c, we show $n/2$ for both small and large flaw densities for comparison.
It is quite clear that the initial decay is faster for the large flaw density and the nonlinear behavior is more evident.
At longer times, however, it may take even longer for the trapped electrons to recombine back to the valence band.
This example illustrates that modeling carrier transients in small-scale silicon waveguides cannot be accurately performed with a single lifetime constant in order to explain our experimental observations.

\begin{figure}
\centering
\includegraphics{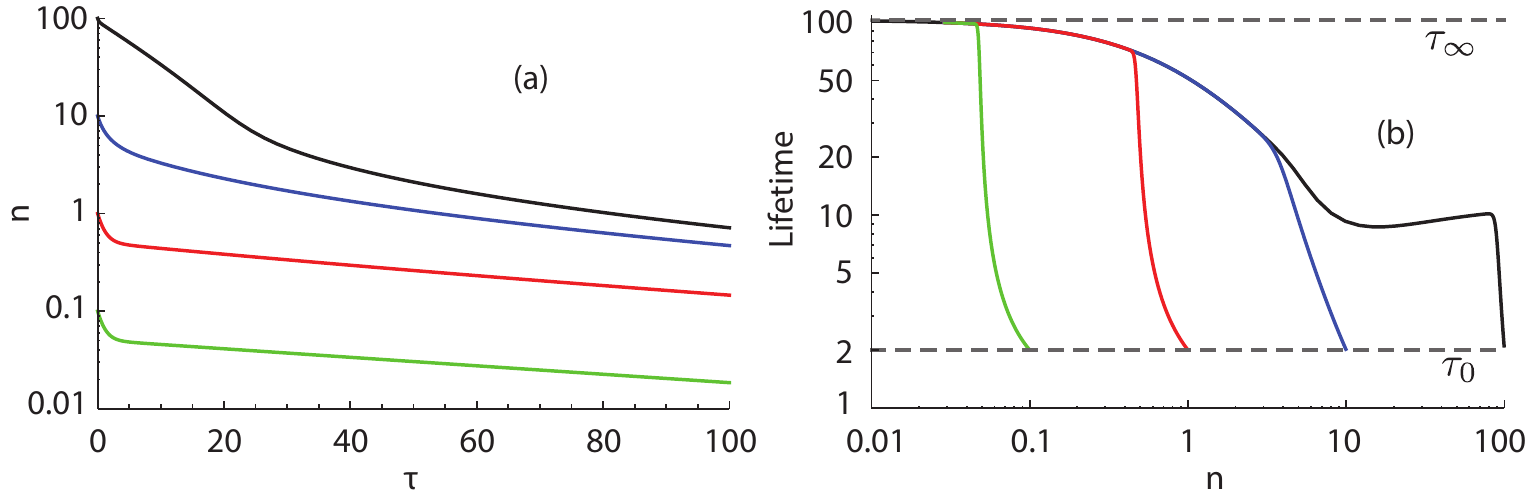}
\caption{(a)~Transient decay and (b)~instantaneous lifetime for different initial carrier densities.
Each curve was calculated using the simulation parameters in Table~\ref{tableparam} and large flaw density ($D=10$).
Each curve in (b) corresponds to the same color curve in (a).}
\label{fig:Nvstime}
\end{figure}

The transient decay dependency on the initial carrier density observed in our experiment can be also qualitatively explained by this model.
Figure~\ref{fig:Nvstime}a shows the transient decay for various initial carrier densities.
All curves were obtained for the same set of parameters as before, assuming relatively large flaw density ($D = 10$), and varying only the initial excess carrier density.
Very clearly, the initial decay is non-exponential and qualitatively agree with the experimental results in Fig.~\ref{Fig:CarrierDynamics}b.
Moreover, none of the curves is a simple time shift of the other.
Take for example the curves corresponding to initial density $n=100$ (in black) and $n=10$ (in blue).
At $\tau \approx 20$, the black curve has reached $n=10$, and the remaining decay curve is completely different than the curve corresponding to initial density $n=10$.
A simple way to evaluate this is to compute the instantaneous carrier lifetime and plot that as a function of the instantaneous carrier density (as performed for the experimental data in Fig.~\ref{Fig:CarrierDynamics}c).
The result is shown in Fig.~\ref{fig:Nvstime}b, where we can see that different curves exhibit different lifetimes for the same instantaneous carrier density.
Once again, a qualitative agreement with the experimental results in Fig.~\ref{Fig:CarrierDynamics}c is obtained.
This memory-like effect on the total density occurs because the ratio between flaw density and initial excess carriers determines how fast the electrons decay at the initial stages of recombination~\cite{ahrenkiel1991intensity}.

The dashed lines in Fig.~\ref{fig:Nvstime}b represent the limits (initial and final) for the normalized carrier lifetime.
These limits can be calculated as $\tau_{0} = 2(1+b)$ and $\tau_{\infty} = D (1+b)\gamma^{-1}[D b + (1+b)^{2}]^{-1}$, respectively~\cite{blakemore2002}.
For flaws near the middle of the bandgap, $b$ is approximately zero, therefore $\tau_{0} = 2$ and $\tau_{\infty} = D\gamma^{-1}$.
Using the definitions in Table~\ref{tableparam}, we can convert the normalized lifetimes to absolute values: $t_0 = \frac{W}{v_{n} \sigma_n D_s}$ and $t_{\infty} = \frac{1}{v_{p} \sigma_p p_0 }$.
Note that, while $t_{0}$ can be reduced by simply decreasing the waveguide dimension $W$, the long lifetime limit $t_{\infty}$ remains unchanged.
From our measurements, $t_{\infty} \approx \SI{300}{ns}$ and, using the parameters from Table~\ref{tableparam}, we can estimate the capture cross-section for holes to be $\sigma_p \approx \SI{2.6e-16}{cm^{2}}$.
This value is in agreement with measurements based on small pulse Deep Level Transient Spectroscopy~\cite{aberle1992impact}.
From our measurements, we have that $t_{0} = \SI{0.8}{ns}$ and, assuming again that $\sigma_n = 8 \sigma_p$, we can estimate the order of magnitude of the surface flaw density as $ D_s = \frac{W}{v_{n} \sigma_n t_0}$.
Using the parameters from Table~\ref{tableparam} we obtain $D_s \approx \SI{1.6e12}{cm^{-2}}$, which is also in agreement with measurements on SiO$_{2}$-Si interface~\cite{aberle1992impact, caplan1979esr}.

The general behavior in the transient decay discussed in the present analysis is relatively robust to the choice of parameters ($a$ and $b$).
One important point is regarding the parameter $b$, which is related to the thermal equilibrium flaw occupancy: as long as the flaw energy level is a few $k_{B} T$ above the Fermi level, its occupancy level is near zero and $b \approx 0$.
For a p-type semiconductor, in which the Fermi level is close to the valence band (in our case, we assumed \SI{0.24}{eV} above the valence band), this practically means the flaw-energy levels can be anywhere near the mid-gap or on the top-part of the bandgap and $b$ is still approximately zero.
This is especially important considering that a continuum of flaw energy states exits in a SiO$_{2}$-Si interface~\cite{aberle1992impact}.
Obviously, the exact decay transient for a continuum of flaw-energies needs to be analyzed in detail.
Other important parameters are the capture cross-sections and flaw densities, which depend on the particular oxide used in the cladding and processing conditions.
Detailed characterization of these parameters would be required for a complete understanding of the decay dynamics.

A final comment is that the assumption $n_{e}(0) = p_{e}(0)$ is valid for an impulse optical excitation.
However, if we use long enough pulses as excitation, then excess electron and hole densities at the beginning of the transient decay, i.e.\ after the pulse, may already be different.
This should be modeled using the complete equations~\ref{eq:dxdt} and~\ref{eq:dydt}.
In both cases, short and long pulses, the excitation term $g$ will depend on the position along the waveguide, since the pump laser will be attenuated as it propagates.
A natural question that arises is then how uniform the carrier density along the propagation length is.
This analysis is discussed in Supplementary Material B.

\section{Conclusions}
\label{s:conclusion}

In conclusion, we have experimentally characterized the recombination dynamics in strip silicon nano-waveguides and revealed a complex decay dynamics, with lifetime varying as recombination evolves in time.
The results were interpreted in terms of trapping in mid-bandgap surface states.
In particular, the analysis suggests an increase in excess charge trapping as the waveguide surface-to-volume ratio increases.
The carrier recombination dynamics observed in our experiments may impact several nonlinear applications and, along with the theoretical discussion, they provide the basis for a more in-depth treatment of free-carrier dynamics.
Finally, our results suggest that proper engineering of surface flaws (both density and capture cross-sections), as well as device geometrical structure may be used to control the decay of free carriers in silicon photonics applications.

\section*{Acknowledgements}
This work was supported by the S\~ao Paulo Research Foundation (FAPESP) under grants 2008/57857, 2012/50259-2, 2013/20180-3, and 2015/04113-0, by the National Council for Scientific and Technological Development (CNPq), grant 574017/2008-9, and by the Coordination for the Improvement of Higher Education Personnel (CAPES).

\begin{table*}[b]
\caption{Definition of normalized variable and parameter values used in the simulations;}
\vspace{0.1in}
\centering
\begin{adjustbox}{max width=\textwidth}
\label{tableparam}
\centering
\begin{tabular}{ccc}
{\bf Parameters} &{\bf Definition} &{\bf Simulation value}\\
\hline
$x = n_{e}/p_{0}$     & Normalized excess electron density          & -- \\
$y = p_{e}/p_{0}$     & Normalized excess hole density              & -- \\
$D = D_f/p_0$         & Normalized flaw density                     & -- \\
$\tau =t/\tau_{n0} $  & Normalized time                             & --\\
$\tau_{n0}=\frac{1}{D_f\sigma_nv_{n}}$ & Electron decay time constant when all flaws are unoccupied & -- \\
$p_{0}$               & Equilibrium hole concentration              & $10^{15}\,$cm$^{-3}$ \\
$v_{n}$            & Electrons thermal velocity                  & \SI{1.7e7}{cm/s} \\
$v_{p}$            & Holes thermal velocity                      & \SI{1.3e7}{cm/s} \\
$\phi$                & Fermi-level                                 & \SI{0.24}{eV} \\
$\phi_{f}$            & Flaw energy level                           & \SI{0.5}{eV} \\
$E_g$                 & Bandgap                                     & \SI{1.12}{eV} \\
$N_{c}$               & Conduction band effective density of states  & \SI{2.8e19}{\cm^{-3}\eV} \\
$N_{v}$               & Valence band effective density of states  & \SI{1e19}{\cm^{-3}\eV} \\
$a = \frac{N_{c}}{N_{v}} \exp\!\left(\frac{\phi+\phi_{f}-2E_g}{k_{B} T}\right)$ & & \num{1e-6}\\
$b = \exp\!\left(\frac{\phi-\phi_{f}}{k_{B} T}\right)$ & & \num{4e-5}\\
$\gamma = \frac{\sigma_{p} v_{p}}{\sigma_{n}v_{n}}$ & & 0.1 \\
\end{tabular}
\end{adjustbox}
\end{table*}

\newpage

\section*{Supplementary material}

\section*{Carrier diffusion}
\label{app:diffusion}

\begin{figure}[h!]
\centering
\includegraphics[scale = 0.95]
{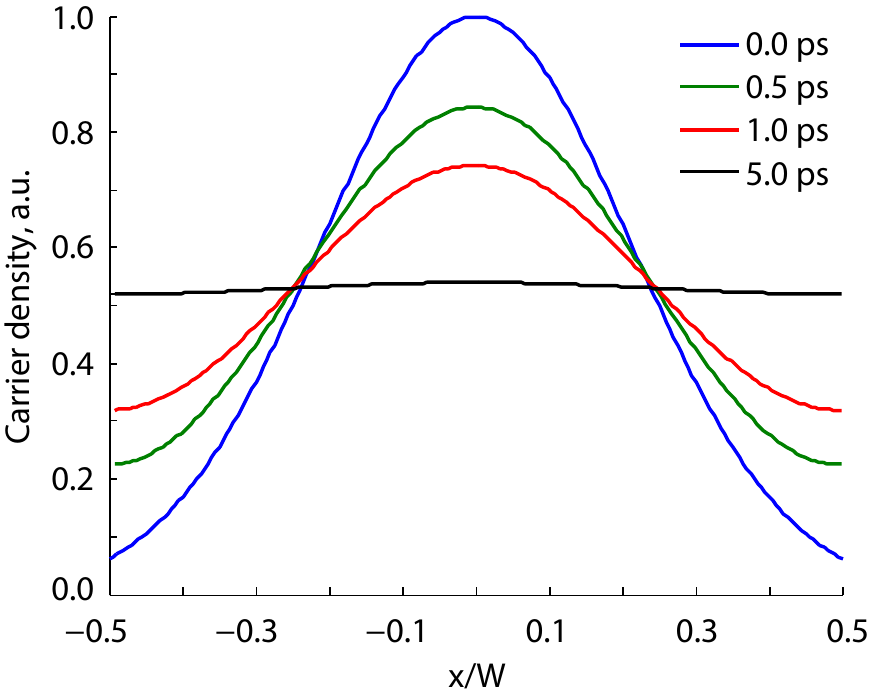}
\caption{Evolution of carrier density over time.}
\label{fig:diffusion}
\end{figure}

Once excess carrier generation has ceased, the spatial carrier distribution diffuses throughout the silicon core until becoming relatively uniform.
The purpose of this analysis is to estimate the time required to reach a uniform carrier distribution.
For that, we assumed an initial Gaussian profile and solved the diffusion equation in one dimension for a waveguide width of $W = \SI{450}{nm}$:
\begin{equation}
\frac{\partial n_i(x,t)}{\partial t}=D_i\frac{\partial^2 n_i(x,t)}{\partial x^2},
\end{equation}
where $n_i$ and $D_i$ represent the carrier density and diffusion coefficient, respectively ($i=e$ for electrons and $i=h$ for holes).
We used $D_e =\SI{3.9e-3}{m^2/s}$ and $D_h = \SI{1.3e-3}{m^2/s}$~\cite{sze2006physics}.
The time evolution for the electron density profile is shown in Fig.~\ref{fig:diffusion} for an initial carrier density $n_e(x,0)= \exp\!\left(-\frac{x^2}{2\sigma^2}\right)$ with $\sigma=0.3W$.
As we can see, the carrier density evolves rapidly towards a uniform distribution in about \SI{5}{ps}, a time scale that is not resolved in the experiments reported in this paper.
For holes, the profile evolves slightly slower, and reaches a uniform distribution in about \SI{15}{ps}.

\section*{Longitudinal dependency}
\label{App:Noverz}

As discussed in the main text, our pump and probe experiment recovers the average carrier density $\bar{N}(t)$ and not the local density $N(z,t)$.
As the pump propagates through the waveguide its intensity drops, thus the generation of excess carrier $N(z,t)$ through {\small TPA} also decreases with distance.
Therefore, it is important to quantify how much the average density $\bar{N}(t)$ deviates from the actual longitudinal distribution $N(z,t)$.
We simulated the propagation of a \SI{130}{ps} pump pulse along with a {\small CW} probe over a \SI{5.9}{mm} long waveguide with pump power varying from \num{0.1} to \SI{1.1}{W} (which reflects the conditions in our experiments).
The dynamic equations for holes and electrons were already presented in the main text (Equations 2 and 3), and the generation mechanism is assumed to be {\small TPA}.
Therefore, the normalized generation rate is given by:
\begin{equation}
g = \frac{\tau_{n0}}{p_0}\frac{\beta I^2}{2 \hbar \omega },
\end{equation}
where $\beta = \SI{0.7}{cm/GW}$ is the {\small TPA} coefficient~\cite{bristow2007}, $I$ is the pump intensity ($I=P/A_{eff}$, with $P$ being the optical power and $A_{eff} = \SI{0.15}{\micro m^2}$ the effective area), and $\omega$ is the pump angular frequency.
Any generation of excess electron-holes by the {\small CW} probe is ignored.
The capture cross-sections for holes and electrons, as well as the flaw density, are those extracted from our experimental data as discussed in the main text: $\sigma_p \approx \SI{2.6e-16}{cm^{-2}}$,  $\sigma_n = 8 \sigma_p$, and $D_s \approx \SI{1.6e12}{cm^{-2}}$.
The propagation equations for the pump intensity, $I$, and probe intensity, $I_P$, are given by:
%
\begin{eqnarray}
\frac{{\rm d}I}{{\rm d}z} &=&-[\alpha + (\sigma_n n_{e} + \sigma_p p_{e}) + \beta I] I,\\
\frac{{\rm d}I_p}{{\rm d}z} &=&-[\alpha + (\sigma_n n_{e} + \sigma_p p_{e}) + 2 \beta I] I_p,
\end{eqnarray}
where $\alpha = \SI{1.4}{dB/cm}$ is the linear attenuation coefficient, while $n_e$ and $p_e$, whose evolution is described by Eq.~2 and 3 of Section~4, are the electron and hole densities, respectively.
The coupled spatio-temporal equations for electrons, holes and for optical intensity were solved using a 1D finite difference method.
The time step was set to \SI{1}{ps} and the spatial step was set to \SI{0.12}{mm} to satisfy the Courant condition.

\begin{figure}
\centering
\includegraphics[scale = 0.8]
{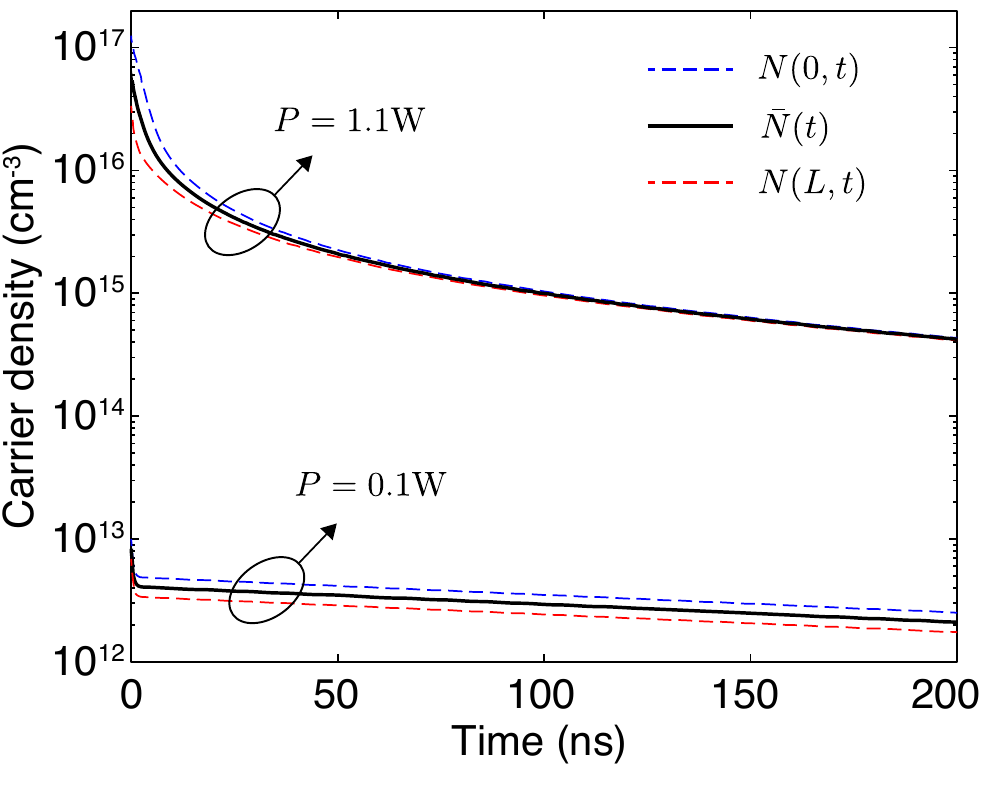}
\caption{Simulated time-resolved carrier densities at $z = 0$, $z = L$ along with the longitudinal average density for pump powers of \SI{1.1}{W} and \SI{0.1}{W}.}
\label{fig:carrierdensity}
\end{figure}

\begin{figure}
\centering
\includegraphics[scale = 0.8]
{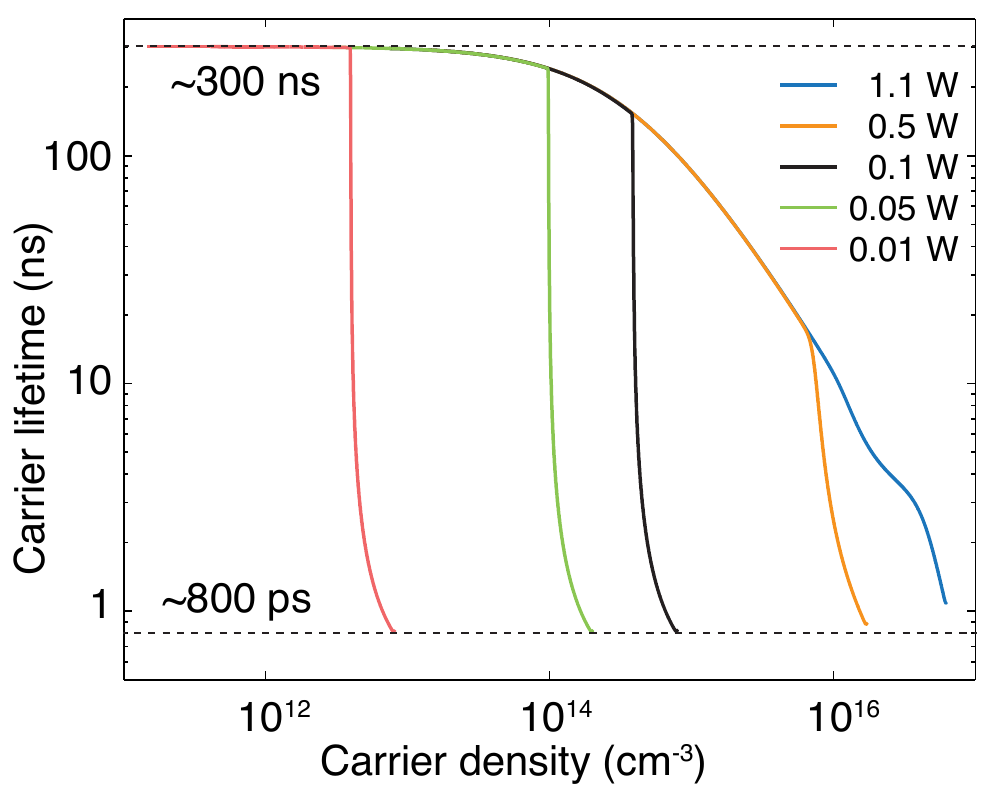}
\caption{Simulated carrier lifetime as a function of carrier density for different pump powers.}
\label{fig:lifetimesim}
\end{figure}

Figure~\ref{fig:carrierdensity} shows the excess carrier density time decay for pump powers of \num{0.1} and \SI{1.1}{W}.
The quantity plotted is $N = (\sigma_n n_{e} + \sigma_p p_{e})/(\sigma_n + \sigma_p)$, as it reflects the carrier density extracted experimentally.
For each power level, the solid curve represents the average density $\bar N(t)$ while the dashed curves represent the maximum $N(0, t)$ and minimum $N(L, t)$ local densities in blue and red colors respectively.
As one can see, the difference between average and local is relatively small, considering the various orders of magnitude spanned by our measurements.
As an example, for the highest peak power (\SI{1.1}{W}) and at the start of the decay transient, the minimum and maximum densities are \SI{4.0e16}{cm^{-3}} and \SI{1.0e17}{cm^{-3}}, while the average is \SI{6.5e16}{cm^{-3}}.
The carrier lifetime curves extracted from our simulation for several pump powers are shown in Fig.~\ref{fig:lifetimesim}.
The general behavior discussed in the main text is not altered by the longitudinal analysis included in this section.

\section*{Numerical derivative}
\label{App:NumDer}

The time-resolved carrier lifetime, $\tau_c(t)$, can be extracted from the time-varying carrier density, $\bar{N}(t)$, as follows~\cite{nagel1999,yablonovitch1986}:
\begin{equation}
\tau_c(t)=-\frac{\bar{N}(t)}{{\rm d}\bar{N}(t)/{\rm d}t},
\end{equation}
which requires computing the time derivative of $\bar{N}(t)$.
We note however that in order to properly measure lifetimes ranging from hundreds of picoseconds to hundreds of nanoseconds, a large bandwidth (\SI{28}{GHz}) photodetector was used, leading to relatively large electrical noise in the detected signal (especially for low carrier density).
Therefore, care must be taken in computing the numerical derivative.
We first averaged a large number of curves.
For example, for the highest pump power (\SI{1.1}{W}), we averaged 25 curves while for the lowest (\SI{70}{mW}) we averaged 400 curves.

\begin{figure}[h!]
\centering
\includegraphics[scale = 0.79]
{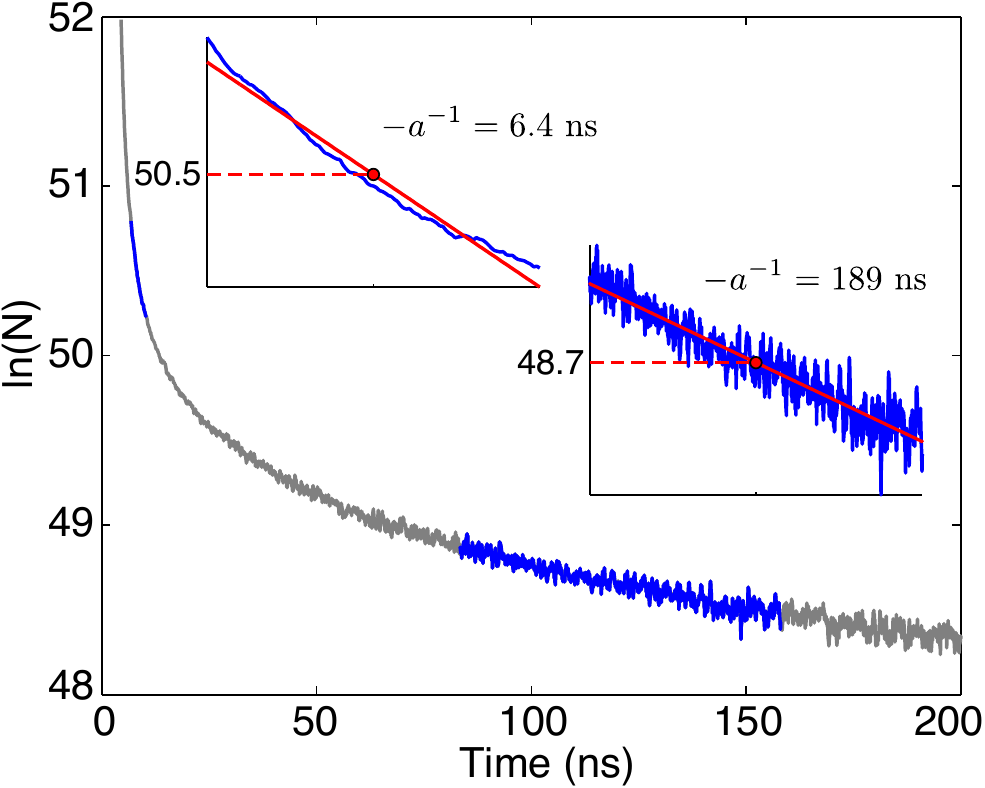}
\caption{Example windowing used in computing the time-resolved carrier lifetime.
The two insets show the lifetime obtained by fitting a linear curve in two different windows, one for high and the other for low carrier densities.}
\label{fig:derivative}
\end{figure}

Then the lifetime was obtained by performing linear regression of the carrier density (in natural logarithmic scale) in windows, as illustrated in Fig.~\ref{fig:derivative}.
Fitting $\ln N(t) |_{\text{fit}}=at+b$ to the measured data in each window, the carrier lifetime can be obtained as $\tau_c=-a^{-1}$.
The size of the window is an important parameter to ensure an accurate calculation of the carrier lifetime: a very wide window cannot follow the change in slope, whereas an excessively short window will result in a lifetime subject to a significant error.
For that reason, we used an adaptive windowing approach, in which higher carrier densities are processed using a shorter window and lower carrier densities are processed using longer ones.

In Fig.~\ref{fig:derivative} the time-resolved carrier density is presented in logarithmic scale alongside with two insets representing the processing of the curve for low and high carrier densities.
For each window, the average time, carrier density, and lifetime can then be calculated.
As a consequence, the time resolution of the lifetime measurement depends on the window size and, typically, results in few points.
In order to improve the time resolution of the method, we can overlap adjacent windows.
In our case we use an overlap ratio of 10\%.

\section*{Effect of the probe signal}
\label{App:Probe}

The probe power used in all of our measurements was set to \SI{-12}{dBm}.
In the present appendix, we analyze the effect of probe power on the measured carrier lifetime, primarily to ensure that the probe level used does not influence the lifetime results.

\begin{figure}[!h]
\centering
\includegraphics[scale = 0.8]
{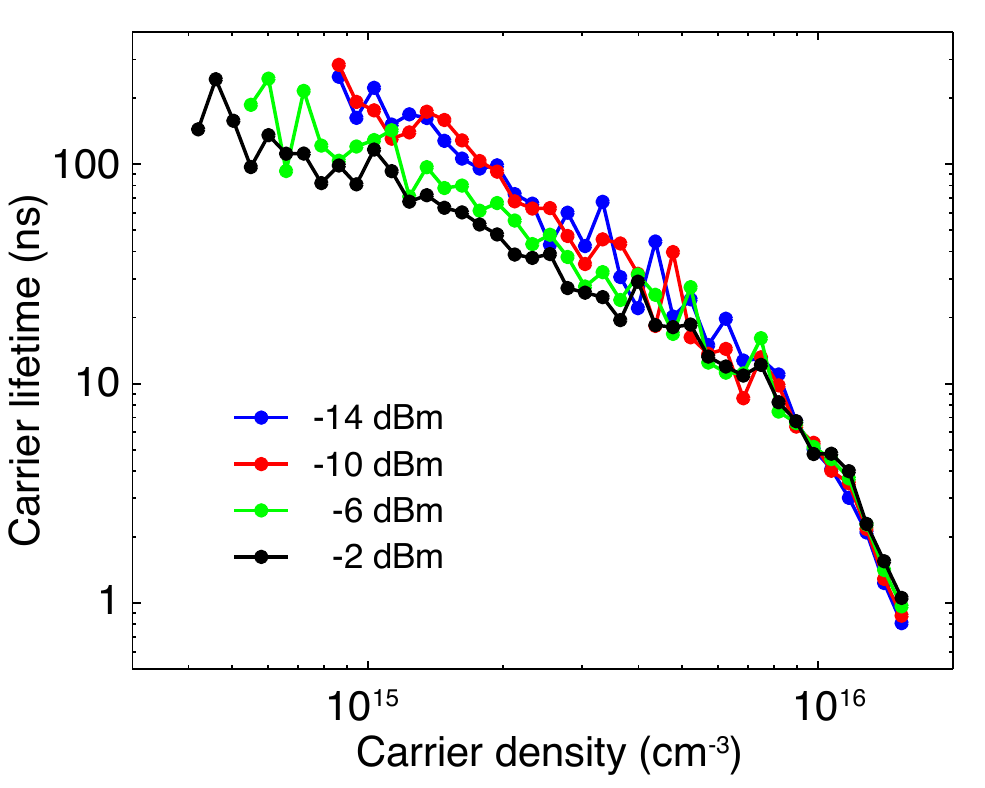}
\caption{Effect of the probe power on the measured carrier lifetime for probe power \num{-14}, \num{-10}, \num{-6} and \SI{-2}{dBm}.
The pump power in this measurement was \SI{280}{mW}.}
\label{fig:probepower}
\end{figure}

Figure~\ref{fig:probepower} shows the measured lifetime in terms of carrier density for probe power ranging from \num{-14} to \SI{-2}{dBm}.
As can be seen, the probe power has little effect for carrier densities above \SI{8e15}{cm^{-3}}.
At lower carrier densities, however, probe power levels at \num{-6} and \SI{-2}{dBm} lead to reduction in the observed lifetime.
For probe power levels of \num{-10} and \SI{-14}{dBm}, there is practically no difference on the lifetime, which justifies our choice of \SI{-12}{dBm}.

Indeed, the fact that at high probe power levels (above \num{-10}~dBm) the measured carrier lifetime for low carrier density is reduced is in agreement with previous experiments in silicon-based solar cells~\cite{macdonald2001}. In our case, the probe signal plays the role of background illumination in the measurements of solar cells.   

\bibliographystyle{ieeetr}
\bibliography{sample}
\end{document}